\documentclass[12pt,leqno]{article}

\usepackage{amsmath}
\usepackage{amsthm}
\usepackage{amssymb}
\usepackage[dvips]{graphicx,color}

\theoremstyle{plain}
\newtheorem{theorem}{Theorem}[section]
\newtheorem{lemma}[theorem]{Lemma}
\newtheorem{proposition}[theorem]{Proposition}

\theoremstyle{definition}
\newtheorem{definition}[theorem]{Definition}

\theoremstyle{remark}
\newtheorem{remark}[theorem]{Remark}

\numberwithin{equation}{section}

\begin{document}

\title{\textbf{Smoothness and monotone decreasingness of the solution to the BCS-Bogoliubov gap equation for superconductivity}}

\author{Shuji Watanabe\\
Division of Mathematical Sciences\\
Graduate School of Engineering, Gunma University\\
4-2 Aramaki-machi, Maebashi 371-8510, Japan\\
Email: shuwatanabe@gunma-u.ac.jp
\and Ken Kuriyama\\
Yamaguchi University\\
2-16-1 Tokiwadai, Ube 755-8611, Japan\\
Email: kuriyama@yamaguchi-u.ac.jp\\}

\date{}

\maketitle

\begin{abstract}
We show the temperature dependence such as smoothness and monotone decreasingness with respect to the temperature of the solution to the BCS-Bogoliubov gap equation for superconductivity. Here the temperature belongs to the closed interval $[0,\, \tau]$ with $\tau>0$ nearly equal to half of the transition temperature. We show that the solution is continuous with respect to both the temperature and the energy, and that the solution is Lipschitz continuous and monotone decreasing with respect to the temperature. Moreover, we show that the solution is partially differentiable with respect to the temperature twice and the second-order partial derivative is continuous with respect to both the temperature and the energy, or that the solution is approximated by such a smooth function.

\medskip

\noindent Mathematics Subject Classification 2010. \    45G10, 47H10, 47N50, 82D55.

\medskip

\noindent Keywords. \    Smoothness, monotone decreasingness, temperature, solution to the BCS-Bogoliubov gap equation, superconductivity.
\end{abstract}

\section{Introduction and main result}
In this paper we show the temperature dependence such as smoothness and monotone decreasingness with respect to the temperature of the solution to the BCS-Bogoliubov gap equation \cite{bcs, bogoliubov} for superconductivity:
\begin{equation}\label{eqn:bcsgapeq}
u(T,\,x)=\int_0^{\hslash\omega_D}
\frac{U(x,\,\xi)\, u(T,\, \xi)}{\,\sqrt{\,\xi^2+u(T,\, \xi)^2\,}\,}\,
\tanh \frac{\,\sqrt{\,\xi^2+u(T,\, \xi)^2\,}\,}{2T}\, d\xi,
\end{equation}
where the solution $u$ is a function of the absolute temperature $T \geq 0$ and the energy $x$
$(0 \leq x \leq \hslash\omega_D)$. The solution $u$ corresponds to the energy gap between the superconducting ground state and the superconducting first excited state, and so the value of the solution is nonnegative, i.e., $u(T,\,x) \geq 0$. The constant $\omega_D>0$ stands for the Debye angular frequency, and the potential $U$ satisfies $U(x,\,\xi)>0$ at all $(x,\,\xi) \in [0, \, \hslash\omega_D]^2$.

In \eqref{eqn:bcsgapeq} we consider the solution $u$ as a function of the absolute temperature $T$ and the energy $x$. Accordingly, we deal with the integral with respect to the energy $\xi$. Sometimes one considers the solution $u$ as a function of the absolute temperature and the wave vector of an electron. Accordingly, instead of the integral with respect to the energy $\xi$ in \eqref{eqn:bcsgapeq}, one deals with the integral with respect to the wave vector over the three dimensional Euclidean space $\mathbb{R}^3$. The existence and uniqueness of the solution to the BCS-Bogoliubov gap equation were established in previous papers \cite{odeh, billardfano, vansevenant, fhns, hhss, haizlseiringer} for each fixed temperature. 
So the temperature dependence such as smoothness and monotone decreasingness with respect to the temperature of the solution is not covered except for the paper \cite{bls}. In \cite{bls} the gap equation in the Hubbard model for a constant potential was studied, and its solution was shown to be strictly decreasing  with respect to the temperature. In this connection, for interdisciplinary reviews of the BCS-Bogoliubov model of superconductivity, see \cite{kuzemsky, kuzemsky2}.

As is well known, studying the temperature dependence of the solution to the BCS-Bogoliubov gap equation is very important in condensed matter physics. This is because studying the temperature dependence of the solution, by dealing with the thermodynamic potential, leads to a mathematical proof of the statement that the transition to the superconducting state is a second-order phase transition in the BCS-Bogoliubov model of superconductivity. In order to give its proof, we have to differentiate the thermodynamic potential, and hence the solution with respect to the temperature twice, and we have to study some properties of the second-order partial derivative of the solution. So it is highly desirable to study the temperature dependence such as smoothness and monotone decreasingness with respect to the temperature of the solution to the BCS-Bogoliubov gap equation \eqref{eqn:bcsgapeq}.

We now define a nonlinear integral operator $A$ by
\begin{equation}\label{eqn:ouroperator}
Au(T,\,x)=\int_0^{\hslash\omega_D}
\frac{U(x,\,\xi)\, u(T,\, \xi)}{\,\sqrt{\,\xi^2+u(T,\, \xi)^2\,}\,}\,
\tanh \frac{\,\sqrt{\,\xi^2+u(T,\, \xi)^2\,}\,}{2T}\, d\xi.
\end{equation}
Here the right side of this equality is exactly the right side of the BCS-Bogoliubov gap equation \eqref{eqn:bcsgapeq}. Since the solution to the BCS-Bogoliubov gap equation is a fixed point of our operator $A$, we apply fixed point theorems to our operator $A$ and study the temperature dependence such as smoothness and monotone decreasingness with respect to the temperature of the solution to the BCS-Bogoliubov gap equation \eqref{eqn:bcsgapeq}.

Let $U_1>0$ is a  positive constant and set $\displaystyle{ U(x,\,\xi)=U_1 }$ at all $(x,\,\xi) \in [0, \, \hslash\omega_D]^2$. Then the solution to the BCS-Bogoliubov gap equation becomes a function of the temperature $T$ only, and so we denote the solution by $\Delta_1: T \mapsto \Delta_1(T)$. Accordingly, the BCS-Bogoliubov gap equation \eqref{eqn:bcsgapeq} is reduced to the simple gap equation \cite{bcs}:
\begin{equation}\label{eqn:delta1}
1=U_1\int_0^{\hslash\omega_D}
 \frac{1}{\,\sqrt{\,\xi^2+\Delta_1(T)^2\,}\,}\,
 \tanh \frac{\, \sqrt{\,\xi^2+\Delta_1(T)^2\,}\,}{2T}\,d\xi, \quad 0 \leq T \leq\tau_1 \, .
\end{equation}
Here the temperature $\tau_1>0$ is defined by (see \cite{bcs})
\[
1=U_1\int_0^{\hslash\omega_D}
\frac{1}{\,\xi\,}\,\tanh \frac{\xi}{\,2\tau_1\,}\,d\xi.
\]
See also Niwa \cite{niwa} and Ziman \cite{ziman}. As is well known in the BCS-Bogoliubov model, physicists and engineers studying superconductivity always assume that there is a unique nonnegative solution $\Delta_1$ to the simple gap equation \eqref{eqn:delta1}, that the solution $\Delta_1$ is continuous and strictly decreasing with respect to the temperature $T$, and that the solution $\Delta_1$ is of class $C^2$ with respect to the temperature $T$, and so on. But, as far as the present authors know, there is no mathematical proof for these assumptions of the BCS-Bogoliubov model. Then, applying the implicit function theorem to the simple gap equation \eqref{eqn:delta1}, one of the present authors obtained the following proposition that indeed gives a mathematical proof for these assumptions mentioned just above:

\begin{proposition}[{\cite[Proposition 1.2]{watanabe-one}}]
Let $U(x,\,\xi)=U_1 \; (>0)$ at all $(x,\,\xi) \in [0, \, \hslash\omega_D]^2$ and set
\[
\Delta=\frac{\, \hslash\omega_D \,}{\,\sinh\frac{1}{\,U_1\,}\,}.
\]
Then there is a unique nonnegative solution $\Delta_1: [\,0,\,\tau_1\,] \to [0,\,\infty)$ to the simple gap equation \eqref{eqn:delta1} such that the solution $\Delta_1$ is continuous and strictly decreasing with respect to the temperature $T$ on the closed interval $[\,0,\,\tau_1\,]$:
\[
\Delta_1(0)=\Delta>\Delta_1(T_1)>\Delta_1(T_2)>\Delta_1(\tau_1)=0, \qquad 0<T_1<T_2<\tau_1.
\]
Moreover, the solution $\Delta_1$ is of class $C^2$ with respect to the temperature $T$ on the interval $[\,0,\,\tau_1\,)$ and satisfies
\[
\Delta_1'(0)=\Delta_1''(0)=0 \quad \mbox{and} \quad \lim_{T\uparrow \tau_1} \Delta_1'(T)=-\infty.
\]
\end{proposition}

\begin{remark}
We set $\Delta_1(T)=0$ for $T>\tau_1$. See figure 1.
\end{remark}

We introduce another positive constant $U_2>0$. Let $0<U_1<U_2$ and set $U(x,\,\xi)=U_2$ at all $(x,\,\xi) \in [0,\, \hslash\omega_D]^2$. Then a similar discussion implies that for $U_2$, there is a unique nonnegative solution $\Delta_2: [\,0,\,\tau_2\,] \to [0,\,\infty)$ to the simple gap equation
\begin{equation}\label{eqn:delta2}
1=U_2\int_0^{\hslash\omega_D}
 \frac{1}{\,\sqrt{\,\xi^2+\Delta_2(T)^2\,}\,}\,
 \tanh \frac{\, \sqrt{\,\xi^2+\Delta_2(T)^2\,}\,}{2T}\,d\xi, \qquad
0\leq T\leq \tau_2.
\end{equation}
Here, $\tau_2>0$ is defined by
\[
1=U_2\int_0^{\hslash\omega_D}
\frac{1}{\,\xi\,}\,\tanh \frac{\xi}{\,2\tau_2\,}\,d\xi.
\]
We again set $\Delta_2(T)=0$ for $T>\tau_2$.

\begin{lemma}[{\cite[Lemma 1.5]{watanabe-one}}] \quad {\rm (a)} The inequality $\tau_1<\tau_2$ holds.

\noindent {\rm (b)} If \   $0\leq T<\tau_2$, then $\Delta_1(T)<\Delta_2(T)$. If \  $T\geq \tau_2$, then $\Delta_1(T)=\Delta_2(T)=0$.
\end{lemma}
See figure 1.  The function $\Delta_2$ has properties similar to those of the function $\Delta_1$.

\begin{figure}[htbp]\hspace{3cm}
\includegraphics[width=8cm]{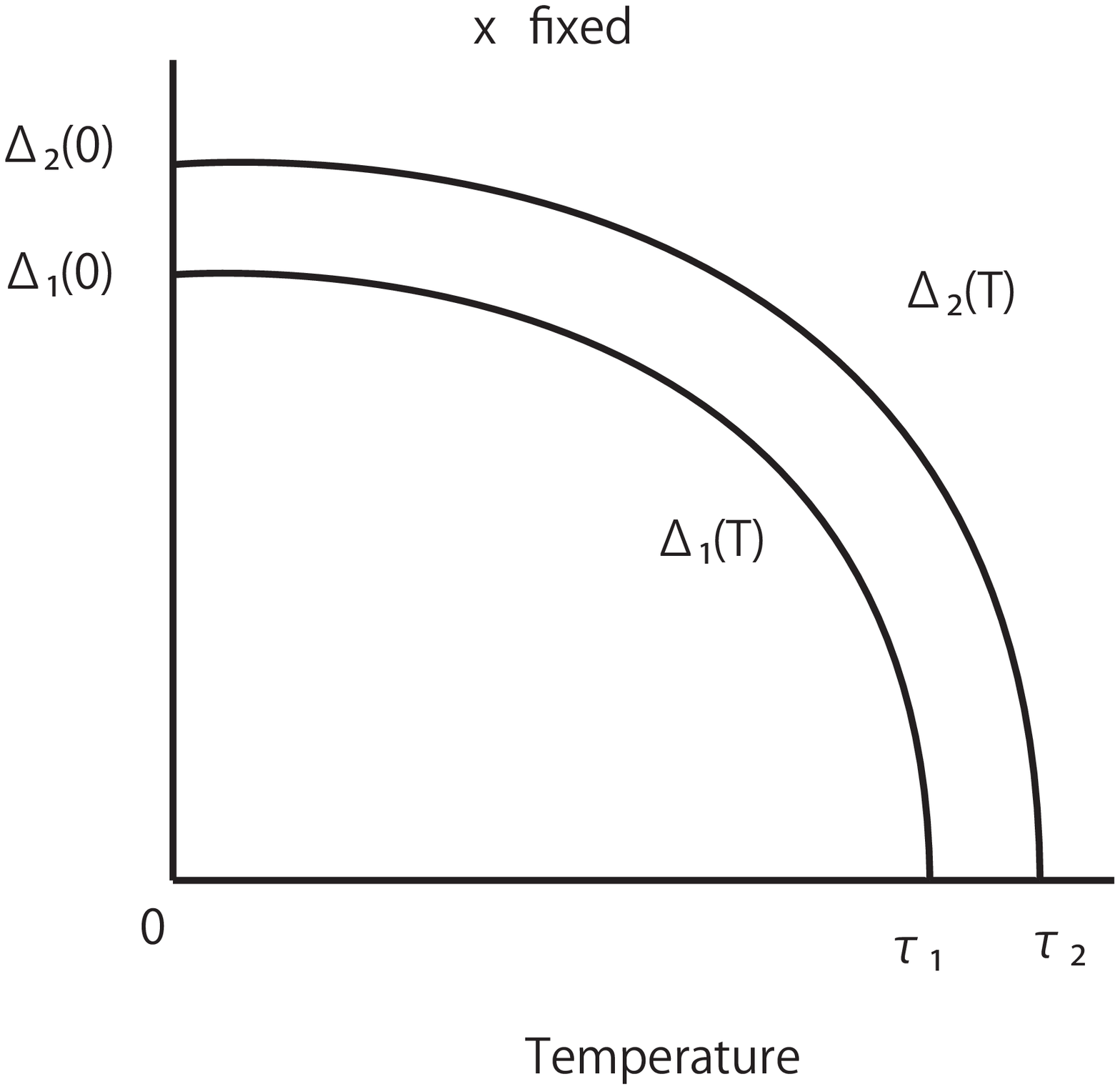}
\caption{\textsf{The graphs of the functions $\Delta_1$ and $\Delta_2$ with the energy $x$ fixed.}}
\end{figure}

We now deal with the BCS-Bogoliubov gap equation \eqref{eqn:bcsgapeq}, where the potential $U$ is not a constant but a function. We assume the following condition on $U$:
\begin{equation}\label{eqn:conditionU}
U(\cdot,\,\cdot) \in C([0,\, \hslash\omega_D]^2), \quad U_1 \leq U(x,\,\xi) \leq U_2 \quad \mbox{at all} \quad (x,\,\xi) \in [0,\, \hslash\omega_D]^2.
\end{equation}
Let $0 \leq T \leq \tau_2$ and fix $T$. We now consider the Banach space $C[0,\, \hslash\omega_D]$ consisting of continuous functions of the energy $x$ only, and deal with the following temperature dependent subset $V_T$:
\[
V_T=\left\{ u(T,\,\cdot) \in C[0,\, \hslash\omega_D]: \; \Delta_1(T) \leq u(T,\,x) \leq \Delta_2(T) \; \mbox{at} \; x \in [0,\, \hslash\omega_D] \right\}.
\]

\begin{remark}
The set $V_T$ depends on the temperature $T$. See figure 1 and 2.
\end{remark}

Applying the Schauder fixed-point theorem to our operator \eqref{eqn:ouroperator} defined on $V_T$, one of the present authors gave another proof of the existence and uniqueness of the nonnegative solution to the BCS-Bogoliubov gap equation \eqref{eqn:bcsgapeq}, which shows how the solution varies with the temperature.
\begin{theorem}[{\cite[Theorem 2.2]{watanabe-one}}] \label{thm:3-1}
Assume \eqref{eqn:conditionU} and fix $T \in [0,\, \tau_2]$. Then there is a unique nonnegative solution $u_0(T,\,\cdot) \in V_T$ to the BCS-Bogoliubov gap equation \eqref{eqn:bcsgapeq}:
\[
u_0(T,\, x)=\int_0^{\hslash\omega_D}
\frac{U(x,\,\xi)\, u_0(T,\, \xi)}{\,\sqrt{\,\xi^2+u_0(T,\, \xi)^2\,}\,}\,
\tanh \frac{\,\sqrt{\,\xi^2+u_0(T,\, \xi)^2\,}\,}{2T}\, d\xi, \, \quad  x \in [0,\, \hslash\omega_D].
\]
Consequently, the solution $u_0(T,\,\cdot)$ with $T$ fixed is continuous with respect to the energy $x$ and varies with the temperature as follows:
\[
\Delta_1(T) \leq u_0(T,\, x) \leq \Delta_2(T) \quad \mbox{at} \quad
(T,\,x) \in [0,\, \tau_2] \times [0,\, \hslash\omega_D].
\]
\end{theorem}

See figure 2.

\begin{figure}[htbp]
\includegraphics[width=18cm]{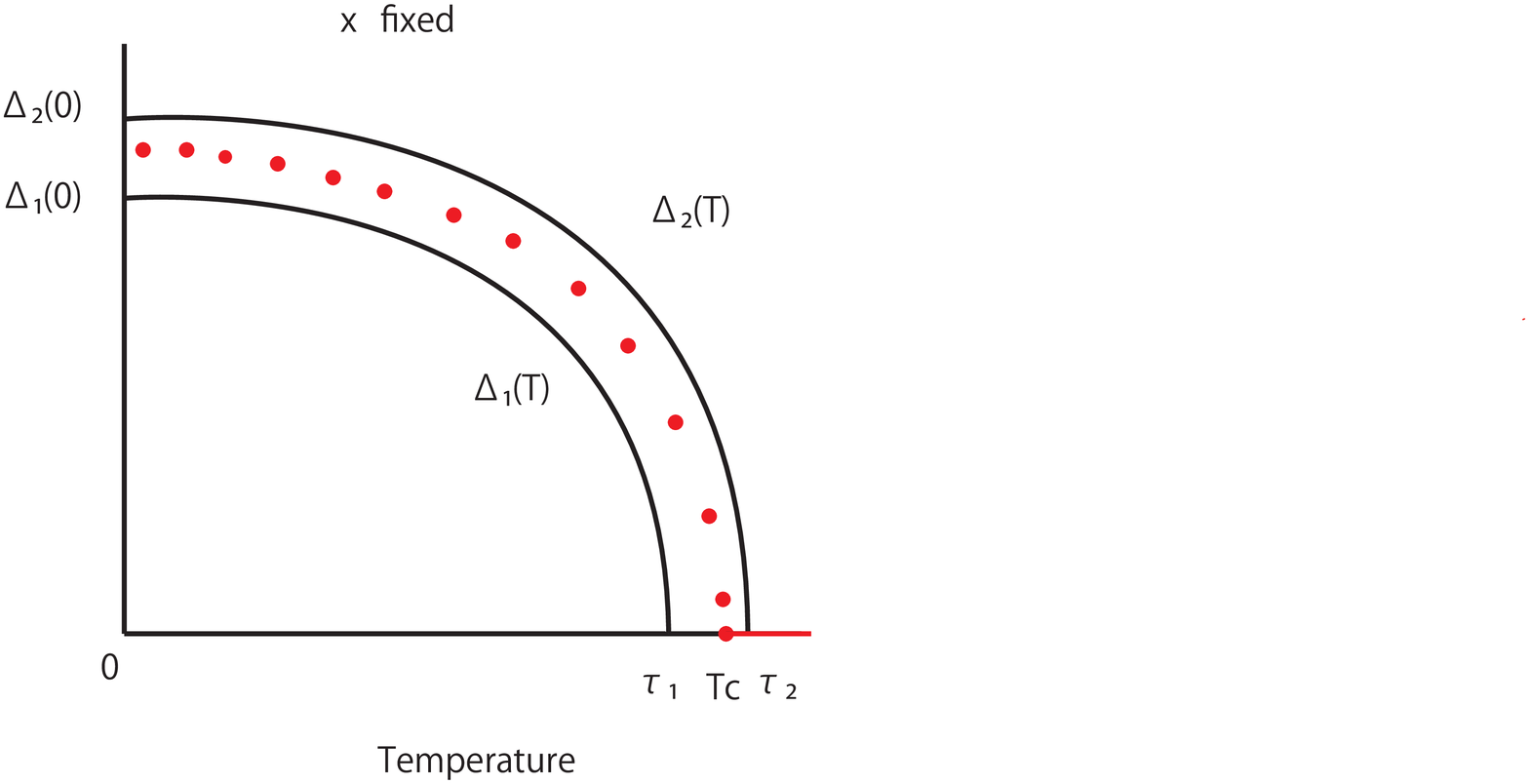}
\caption{\textsf{For each fixed $T$, the solution $u_0(T,\, x)$ lies between $\Delta_1(T)$ and $\Delta_2(T)$.}}
\end{figure}

Superconductivity is observed when the temperature $T$ satisfies $T < T_c$. Here, $T_c$ is the transition temperature (critical temperature) and divides superconductivity $(T<T_c)$ and normal conductivity $(T>T_c)$. The existence and uniqueness of the transition temperature $T_c$ were pointed out in previous papers \cite{fhns, hhss, haizlseiringer, vansevenant}. In our case, we can define it as follows:
\begin{definition}\label{dfn:tc}
Let $u_0(T,\,\cdot)$ be as in Theorem \ref{thm:3-1}. Then the transition temperature $T_c$ is defined by 
\[
T_c=\inf\{ T>0: \, u_0(T,\, x)=0 \quad \mbox{at all} \quad x \in [\varepsilon,\, \hslash\omega_D] \}.
\]
\end{definition}
Note that the transition temperature $T_c$ satisfies $\tau_1\leq T_c \leq \tau_2$. Let $u_0(T,\,\cdot)$ be as in Theorem \ref{thm:3-1}. A straightforward calculation gives that if there is a point $x_1 \in [0,\, \hslash\omega_D]$ satisfying $u_0(T,\, x_1)=0$, then $u_0(T,\, x)=0$ at all $x \in [0,\, \hslash\omega_D]$. We then set $u_0(T,\, x)=0$ at all $x \in [\varepsilon,\, \hslash\omega_D]$ for $T \geq T_c$. We thus see that $u_0(T,\, x)>0$ at all $x \in [0,\, \hslash\omega_D]$ for $0 \leq T <T_c$ and that $u_0(T,\, x)=0$ at all $x \in [0,\, \hslash\omega_D]$ for $T \geq T_c$. See figure 2.

\begin{remark}
Theorem \ref{thm:3-1} tells us nothing about continuity of the solution $u_0$ with respect to the temperature $T$. Applying the Banach fixed-point theorem, we then showed in \cite[Theorem 1.2]{watanabe-two} that the solution $u_0$ is indeed continuous both with respect to the temperature $T$ and with respect to the energy $x$ under the restriction that the temperature $T$ is sufficiently small. See also \cite{watanabe-three}.
\end{remark}

When the potential $U(\cdot,\,\cdot)$ is not a constant but a function, one of the present authors \cite{watanabe-four} studied the temperature dependence such as smoothness and monotone decreasingness of the solution to the BCS-Bogoliubov gap equation \eqref{eqn:bcsgapeq} with respect to the temperature near the transition temperature $T_c$, and gave the behavior of the solution near the transition temperature $T_c$. Then, dealing with the thermodynamic potential, it was shown that the transition to the superconducting state is a second-order phase transition from the viewpoint of operator theory \cite{watanabe-four}. Moreover, the exact and explicit expression for the gap in the specific heat at constant volume at the transition temperature $T_c$ was also obtained in \cite{watanabe-four}.

Let us denote by $z_0 >0$ a unique solution to the equation $\displaystyle{ \frac{2}{\,  z \,} = \tanh z }$
\quad  $(z>0)$. Note that $z_0$ is nearly equal to 2.07 and that $\displaystyle{ \frac{2}{\,  z \,} \leq \tanh z }$ for $z \geq z_0$ . Let  $\tau_0 \, (>0)$ satisfy
\begin{equation}\label{eq:tau0}
\Delta_1(\tau_0) = 2z_0\tau_0 \,.
\end{equation}
From \eqref{eq:tau0} it follows immediately that $(0 <)\, \tau_0 <\tau_1$.

\begin{remark}
Observed values in many experiments by using superconductors imply the temperature $\tau_0$ is nearly equal to $T_c/2$.
\end{remark}

Let $0<\tau < \tau_0$ and fix $\tau$. We then deal with the following subset $V$ of the Banach space $C([0,\, \tau] \times [0,\,\hslash\omega_D])$:
\begin{eqnarray}
V &=& \left\{ u \in C([0,\, \tau] \times [0,\,\hslash\omega_D]) :  0 \leq u(T,\,x)-u(T',\,x) \leq \gamma \left( T'-T \right) \; \; (T<T'),
\right.  \nonumber \\
& & \left. \Delta_1(T) \leq u(T,\,x) \leq \Delta_2(T), \   u \   \mbox{is partially differentiable with respect to} \  T \   \mbox{twice}, 
\right.  \nonumber \\
& & \left. \frac{\,\partial u\,}{\partial T}, \   \frac{\,\partial^2 u\,}{\partial T^2}  \in C([0,\, \tau] \times [0,\,\hslash\omega_D])
\right\}. \nonumber
\end{eqnarray}
Here, $\gamma>0$ is defined by \eqref{eqn:gamma} below. Let us define our operator \eqref{eqn:ouroperator} on the subset $V$ of the Banach space $C([0,\, \tau] \times [0,\,\hslash\omega_D])$. We denote by $\overline{V}$ the closure of the subset $V$ with respect to the norm $\| \cdot \|$ of the Banach space $C([0,\, \tau] \times [0,\,\hslash\omega_D])$. 

\begin{remark}\label{rmk:gamma}
The constant $\gamma>0$ depends neither on $u \in \overline{V}$, nor on $T \in [0,\, \tau]$, nor on $x \in [0,\,\hslash\omega_D]$. See \eqref{eqn:gamma}.
\end{remark}

The following is our main result.

\begin{theorem}\label{thm:main}
Assume \eqref{eqn:conditionU}. Let $\tau$ and $V$ be as above. Then the operator $A:\, \overline{V} \to \overline{V}$ has a unique fixed point $u_0 \in \overline{V}$, and so there is a unique nonnegative solution $u_0 \in \overline{V}$ to the BCS-Bogoliubov gap equation \eqref{eqn:bcsgapeq}:
\[
u_0(T,\,x)=\int_0^{\hslash\omega_D}
\frac{U(x,\,\xi)\, u_0(T,\, \xi)}{\,\sqrt{\,\xi^2+u_0(T,\, \xi)^2\,}\,}\,
\tanh \frac{\,\sqrt{\,\xi^2+u_0(T,\, \xi)^2\,}\,}{2T}\, d\xi, \quad 0 \leq T \leq \tau\, , \quad 0 \leq x \leq \hslash\omega_D \,.
\]
Consequently, the solution $u_0$ is continuous on $[0,\, \tau] \times [0,\,\hslash\omega_D]$, i.e., the solution $u_0$ is continuous with respect to both the temperature $T$ and the energy $x$. Moreover, the solution $u_0$ is Lipschitz continuous and monotone decreasing with respect to the temperature $T$, and satisfies $\Delta_1(T) \leq u_0(T,\,x) \leq \Delta_2(T)$ at all $(T,\,x) \in [0,\, \tau] \times [0,\,\hslash\omega_D]$. Furthermore, if $u_0 \in V$, then the solution $u_0$ is partially differentiable with respect to the temperature $T$ twice and the second-order partial derivative is continuous with respect to both the temperature $T$ and the energy $x$. On the other hand, if $u_0 \in \overline{V} \setminus V$, then the solution $u_0$ is approximated by such a smooth element of the subset $V$ with respect to the norm $\| \cdot \|$ of the Banach space $C([0,\, \tau] \times [0,\,\hslash\omega_D])$.
\end{theorem}

See figure 3 for the graph of the solution $u_0$ with the energy $x$ fixed.

\begin{figure}[htbp]\hspace{-2.6cm}
\includegraphics[width=17cm]{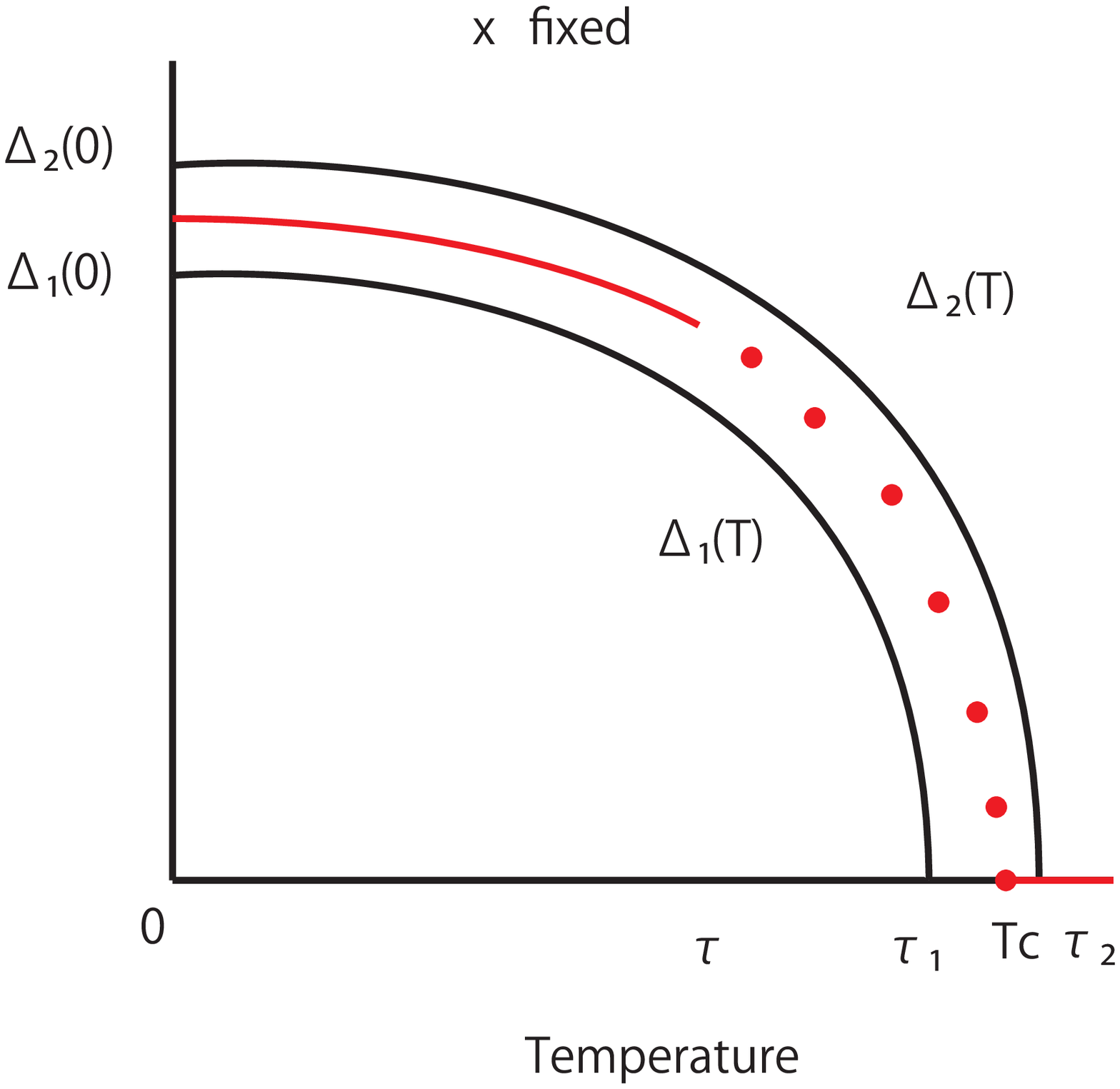}
\caption{\textsf{The solution $u_0$ belongs to the subset  $\overline{V}$.}}
\end{figure}

\section{Proof of Theorem \ref{thm:main}}

We prove Theorem \ref{thm:main} in a sequence of lemmas.

\begin{lemma}\label{lm:functionF}
Let $0<\tau < \tau_0$ and fix $\tau$. Define a function $F$ on $[0, \, \tau]$ by
\[
F(T)=\int_0^{\hslash\omega_D} \frac{1}{\,\sqrt{\,\xi^2+\Delta_1(T)^2\,}\,}\,
 \tanh \frac{\, \sqrt{\,\xi^2+\Delta_1(T)^2\,}\,}{2\tau_0}\,d\xi, \qquad T \in [0,\, \tau].
\]
Then the function $F$ is continuous on $[0, \, \tau]$.
\end{lemma}

\begin{proof}
Let $T \in [0,\, \tau]$. Note that $\displaystyle{ \frac{z}{\, \cosh^2z \,} \leq \tanh z }$ \  $(z \geq 0)$ and that $\displaystyle{ \frac{\,\tanh z \,}{z} \leq 1 }$ \  $(z \geq 0)$. Then
\begin{eqnarray*}
\, & & \left| F(T+h)-F(T) \right| \nonumber \\
&\leq& \int_0^{\hslash\omega_D}
 \frac{\, \left| \Delta_1(T+h)^2-\Delta_1(T)^2 \right| \,}{\,2\left( \,\xi^2+d \,\right)^{3/2}\,}\, \left\{
 \tanh \frac{\, \sqrt{\,\xi^2+d \,}\,}{2\tau_0}+\frac{\, \sqrt{\, \xi^2+d\,}\,}{2\tau_0}
\frac{1}{\, \cosh^2 \frac{\, \sqrt{\,\xi^2+d \,}\,}{2\tau_0} \,} \right\}
\,d\xi  \nonumber \\
&\leq& \left| \Delta_1(T+h)^2-\Delta_1(T)^2 \right| \, \int_0^{\hslash\omega_D}
 \frac{1}{\,\left( \,\xi^2+d \,\right)^{3/2}\,}\, 
 \tanh \frac{\, \sqrt{\,\xi^2+d \,}\,}{2\tau_0}\,d\xi  \nonumber \\
&\leq& \left| \Delta_1(T+h)^2-\Delta_1(T)^2 \right| \, \int_0^{\hslash\omega_D}
 \frac{d\xi}{\,2\tau_0\left( \,\xi^2+d \,\right)\,} \, . \nonumber \\
\end{eqnarray*}
Here, $d$ is between $\Delta_1(T+h)^2$ and $\Delta_1(T)^2$. Since $d\geq \Delta_1(\tau)^2$, it follows that
\[
\left| F(T+h)-F(T) \right| \leq \left| \Delta_1(T+h)^2-\Delta_1(T)^2 \right|
 \frac{1}{\, 2\tau_0\Delta_1(\tau) \,} \, \arctan \frac{\, \hslash\omega_D \,}{\Delta_1(\tau)}.
\]
Continuity of the function $\Delta_1$ proves the lemma.
\end{proof}

Let $0<\tau < \tau_0$ and fix $\tau$. In view of Lemma \ref{lm:functionF}, we set
\begin{eqnarray}
a&=&\max_{0 \leq T \leq \tau} \int_0^{\hslash\omega_D}
 \frac{1}{\,\sqrt{\,\xi^2+\Delta_1(T)^2\,}\,}\,
 \tanh \frac{\, \sqrt{\,\xi^2+\Delta_1(T)^2\,}\,}{2\tau_0}\,d\xi, \label{eq:a} \\
b&=&\frac{32\tau^2}{\, \Delta_1(\tau)^2 \,} \, \arctan \frac{\, \hslash\omega_D \,}{\Delta_1(\tau)}.
\nonumber
\end{eqnarray}
Then, for $T \in [0,\, \tau]$, 
\begin{eqnarray*}
1&=&U_1\int_0^{\hslash\omega_D}
 \frac{1}{\,\sqrt{\,\xi^2+\Delta_1(T)^2\,}\,}\,
 \tanh \frac{\, \sqrt{\,\xi^2+\Delta_1(T)^2\,}\,}{2T}\,d\xi \nonumber \\
&>&U_1\int_0^{\hslash\omega_D}
 \frac{1}{\,\sqrt{\,\xi^2+\Delta_1(T)^2\,}\,}\,
 \tanh \frac{\, \sqrt{\,\xi^2+\Delta_1(T)^2\,}\,}{2\tau_0}\,d\xi \nonumber \\
\end{eqnarray*}
by \eqref{eqn:delta1}. Lemma \ref{lm:functionF} implies $1>U_1a$, where $a$ is that in
\eqref{eq:a}. We choose $U_2 \, (>U_1)$ such that $1>U_2a$ holds true. Set
\begin{equation}\label{eqn:gamma}
\gamma = \frac{\, U_2b\,}{\, 1-U_2a \,} \quad (>0).
\end{equation}
As mentioned in Remark \ref{rmk:gamma}, the constant $\gamma>0$ depends neither on $u \in \overline{V}$, nor on $T \in [0,\, \tau]$, nor on $x \in [0,\,\hslash\omega_D]$.

\begin{lemma}\label{lm:G}
Let $T \in [0,\,\tau_0]$ and let $X \in [\Delta_1(\tau_0)^2,\,\infty)$. Define a function $G$ by
\[
G(T,\, X,\,\xi)=\xi^2\tanh \frac{\, \sqrt{\,\xi^2+X \,}\,}{2T}+\frac{4XT}{\, \sqrt{\,\xi^2+X \,}\,}, \quad
0 \leq \xi \leq \hslash\omega_D\, .
\]
Then $G$ is a monotone increasing function with respect to $T \in [0,\,\tau_0]$. Consequently, $G(T,\, X,\,\xi) \leq G(\tau_0,\, X,\,\xi)$.
\end{lemma}

\begin{proof}
A straightforward calculation gives
\[
\frac{\, \partial G \,}{\partial T}=\frac{1}{\,  2\sqrt{\xi^2+X}\,}
\left(  \sqrt{8X}+\frac{\xi\sqrt{\xi^2+X}}{\,  T\cosh \frac{\sqrt{\xi^2+X}}{2T}\,}  \right)
\left(  \sqrt{8X}-\frac{\xi\sqrt{\xi^2+X}}{\,  T\cosh \frac{\sqrt{\xi^2+X}}{2T}\,}  \right).
\]
Since $\displaystyle{ \frac{z}{\, \cosh z \,} \leq \frac{2}{\, z\,} }$ \  $(z \geq 0)$, it follows from
\eqref{eq:tau0} that
\begin{eqnarray*}
\sqrt{8X}-\frac{\xi\sqrt{\xi^2+X}}{\,  T\cosh \frac{\sqrt{\xi^2+X}}{2T}\,} &\geq&
\sqrt{8X}-\frac{8\xi T}{\,  \sqrt{\xi^2+X} \,} \nonumber \\
&=& \frac{\sqrt{8X}\sqrt{\xi^2+X}-8\xi T}{\,  \sqrt{\xi^2+X} \,} \nonumber \\
&\geq& \frac{\sqrt{8}\, \Delta_1(\tau_0)\xi-8\xi \tau_0}{\,  \sqrt{\xi^2+X} \,} \nonumber \\
&=& \frac{2\sqrt{8}\xi \tau_0}{\,  \sqrt{\xi^2+X} \,}\left( z_0-\sqrt{2} \right) \nonumber \\
&\geq& 0.
\end{eqnarray*}
Note that $\sqrt{X} \geq \Delta_1(\tau_0)$ and that $z_0$ is nearly equal to 2.07. The result thus follows.
\end{proof}

A straightforward calculation gives the following.
\begin{lemma}\label{lm:setvvar}
The subset $\overline{V}$ is bounded, closed, convex and nonempty.
\end{lemma}

\begin{lemma}\label{lm:delta}
If $u \in V$, then $\Delta_1(T) \leq Au(T,\,x) \leq \Delta_2(T)$ at all $(T,\,x) \in [0,\, \tau] \times [0,\,\hslash\omega_D]$.
\end{lemma}

\begin{proof}
Since $u(T,\,x) \leq \Delta_2(T)$, it follows that
\[
\frac{ u(T,\, \xi) }{ \,\sqrt{\,\xi^2+u(T,\, \xi)^2\, }\, } \leq
\frac{ \Delta_2(T) }{ \,\sqrt{\,\xi^2+ \Delta_2(T)^2\, }\, }.
\]
Therefore \eqref{eqn:delta2} gives
\[
Au(T,\,x) \leq U_2\int_0^{\hslash\omega_D} \frac{ \Delta_2(T) }{\,\sqrt{\,\xi^2+\Delta_2(T)^2\,}\,}\,
 \tanh \frac{\, \sqrt{\,\xi^2+\Delta_2(T)^2\,}\,}{2T}\,d\xi = \Delta_2(T).
\]
Similarly we can show the rest.
\end{proof}

\begin{lemma}\label{lm:gamma}
For $T, \, T' \in [0,\, \tau]$, let $T<T'$. If $u \in V$, then
\[
0 \leq Au(T,\,x)-Au(T',\,x) \leq \gamma \left( T'-T \right), \qquad x \in [0,\,\hslash\omega_D].
\]
\end{lemma}

\begin{proof} \   Step 1. We first show $Au(T,\,x)-Au(T',\,x) \geq 0$. 
\[
Au(T,\,x)-Au(T',\,x)=\int_0^{\hslash\omega_D} U(x,\,\xi) \left( K_1 + K_2 \right) \, d\xi,
\]
where
\begin{eqnarray}
K_1 &=& \frac{u(T,\, \xi)}{\,\sqrt{\,\xi^2+u(T,\, \xi)^2\,}\,}\tanh \frac{\,\sqrt{\,\xi^2+u(T,\, \xi)^2\,}\,}{2T}
-\frac{u(T',\, \xi)}{\,\sqrt{\,\xi^2+u(T',\, \xi)^2\,}\,}\tanh \frac{\,\sqrt{\,\xi^2+u(T',\, \xi)^2\,}\,}{2T} \,,
\nonumber \\
K_2 &=& \frac{u(T',\, \xi)}{\,\sqrt{\,\xi^2+u(T',\, \xi)^2\,}\,}\left\{
\tanh \frac{\,\sqrt{\,\xi^2+u(T',\, \xi)^2\,}\,}{2T}
-\tanh \frac{\,\sqrt{\,\xi^2+u(T',\, \xi)^2\,}\,}{2T'} \right\}\,.  \nonumber
\end{eqnarray}
Since $u(T,\,\xi) \geq u(T',\,\xi)$, it follows that
\[
\frac{u(T,\, \xi)}{\,\sqrt{\,\xi^2+u(T,\, \xi)^2\,}\,} \geq \frac{u(T',\, \xi)}{\,\sqrt{\,\xi^2+u(T',\, \xi)^2\,}\,}.
\]
Hence $K_1\geq 0$. Clearly, $K_2\geq 0$. Thus $Au(T,\,x)-Au(T',\,x)\geq 0$.

Step 2. \   We next show $Au(T,\,x)-Au(T',\,x) \leq \gamma \left( T'-T \right)$. \   Since $\displaystyle{ \frac{z}{\,\cosh^2z\,} \leq \frac{2}{\,  z\,}}$ \   $(z \geq 0)$, it follows from Lemma \ref{lm:G} that
\begin{eqnarray*}
K_1&=&\frac{1}{\,\left( \xi^2+c^2 \right)^{3/2} \,}\left\{ \xi^2\tanh
\frac{\,\sqrt{\,\xi^2+c^2\,}\,}{2T} + \frac{\, c^2\, \sqrt{\,\xi^2+c^2\,}\,}{\, 2T\,\cosh^2 \frac{\,\sqrt{\,\xi^2+c^2\,}\,}{2T}\,} \right\} \left\{ u(T,\,\xi)-u(T',\,\xi) \right\} \nonumber \\
&\leq& \frac{1}{\,\left( \xi^2+c^2 \right)^{3/2} \,} \, G(T,\, c^2,\, \xi) \, \gamma (T'-T) \nonumber \\
&\leq& \frac{1}{\,\left( \xi^2+c^2 \right)^{3/2} \,} \, G(\tau_0,\, c^2,\, \xi) 
\, \gamma (T'-T), \nonumber \\
\end{eqnarray*}
where $c$ satisfies $u(T,\,\xi) >c>u(T',\,\xi)$ and depends on $T$, $T'$, $\xi$ and $u$. Note that
\[
\frac{\,\sqrt{\,\xi^2+c^2\,}\,}{2\tau_0} \geq \frac{\,\sqrt{\, c^2\,}\,}{2\tau_0} >
\frac{\,\Delta_1(\tau_0)\,}{2\tau_0}=z_0
\]
by \eqref{eq:tau0}. The substitution $\displaystyle{ z=\frac{\,\sqrt{\,\xi^2+c^2\,}\,}{2\tau_0} }$ therefore turns \   $\displaystyle{ \frac{2}{\,  z \,} \leq \tanh z }$ \    $(z \geq z_0)$ \   into
\[
\frac{4\tau_0}{\,\sqrt{\,\xi^2+c^2\,}\,} \leq \tanh \frac{\,\sqrt{\,\xi^2+c^2\,}\,}{2\tau_0}.
\]
Hence
\begin{eqnarray*}
K_1 &\leq& \frac{1}{\,\sqrt{\,\xi^2+c^2\,}\,} \tanh \frac{\,\sqrt{\,\xi^2+c^2\,}\,}{2\tau_0}
\, \gamma (T'-T) \nonumber \\
&\leq& \frac{1}{\,\sqrt{\,\xi^2+\Delta_1(T')^2\,}\,} \tanh \frac{\,\sqrt{\,\xi^2+\Delta_1(T')^2\,}\,}{2\tau_0} \, \gamma (T'-T). \nonumber
\end{eqnarray*}
Since 
$\displaystyle{ \frac{z}{\,\cosh z\,} \leq \frac{\,2\,}{z}}$ \   $(z \geq 0)$, it follows that
\begin{eqnarray*}
K_2 &=& \frac{ \, 2u(T',\,\xi)(T'-T) \,}{ \xi^2+u(T',\,\xi)^2 } \left\{
\frac{\,\sqrt{\,\xi^2+u(T',\,\xi)^2\,}\,}{2T''} \frac{1}{\, \cosh \frac{\,\sqrt{\,\xi^2+u(T',\,\xi)^2\,}\,}{2T''}  \, } 
 \right\}^2  \nonumber \\
&\leq& \frac{ \, 2u(T',\,\xi)(T'-T) \,}{ \xi^2+u(T',\,\xi)^2 }\frac{16(T'')^2}{\, \xi^2+u(T',\,\xi)^2} 
 \nonumber \\
&\leq& \frac{ \,  (T'-T)\,  32\tau^2\,  }{ \Delta_1(\tau)  }\frac{1}{\, \xi^2+\Delta_1(\tau)^2},
\end{eqnarray*}
where $T<T''<T'$. Thus, by \eqref{eqn:gamma},
\begin{eqnarray*}
& & Au(T,\,x)-Au(T',\,x) \nonumber \\
&\leq& (T'-T) \, U_2\int_0^{\hslash\omega_D} \left( 
\frac{\gamma}{\,\sqrt{\,\xi^2+\Delta_1(T')^2\,}\,}
\tanh \frac{\,\sqrt{\,\xi^2+\Delta_1(T')^2\,}\,}{2\tau_0} 
+\frac{ \,  32\tau^2\,  }{ \, \Delta_1(\tau) \,  }\frac{1}{\, \xi^2+\Delta_1(\tau)^2}
\right) \, d\xi \nonumber \\
&\leq&  (T'-T) \, U_2 \left( \gamma \, a+b  \right) \nonumber \\
&=&  \gamma \, (T'-T). \nonumber \\
\end{eqnarray*}
\end{proof}

\begin{lemma}\label{lm:contin}
If $u \in V$, then $Au \in C([0,\, \tau] \times [0,\,\hslash\omega_D])$.
\end{lemma}

\begin{proof}
Let $T<T'$. Then
\begin{equation}\label{eqn:au}
\left| Au(T,\,x)-Au(T',\,x') \right| \leq \left| Au(T,\,x)-Au(T',\,x) \right| +
\left| Au(T',\,x)-Au(T',\,x') \right|.
\end{equation}
Since $U(\cdot, \,\cdot)$ is uniformly continuous, for an arbitrary $\varepsilon>0$, there is a $\delta_1>0$ such that
$|x-x'|<\delta_1$ implies
\[
\left| U(x,\, \xi)-U(x',\, \xi) \right| <\frac{\varepsilon}{\, 2\hslash\omega_D \,}.
\]
Note that the $\delta_1>0$ depends neither on $x$, nor on $x'$, nor on $\xi$, nor on $u \in V$. Hence the second term on the right of \eqref{eqn:au} becomes
\[
\left| Au(T',\,x)-Au(T',\,x') \right| \leq \int_0^{\hslash\omega_D} \left| U(x,\, \xi)-U(x',\, \xi) \right| \,d\xi
<\frac{\,\varepsilon\,}{2}.
\]
On the other hand, the first term on the right of \eqref{eqn:au} becomes
\[
\left| Au(T,\,x)-Au(T',\,x) \right| \leq \gamma (T'-T)<\frac{\,\varepsilon\,}{2}
\]
by the preceding lemma. Here, $T'-T<\varepsilon/(2\gamma)$. Thus
\[
\left| Au(T,\,x)-Au(T',\,x') \right|<\varepsilon, \quad
 (T'-T)+\left| x-x' \right|<\delta=\min\left( \delta_1,\, \frac{\varepsilon}{\,  2\gamma \,} \right).
\]
Note that the $\delta>0$ depends neither on $x$, nor on $x'$, nor on $\xi$, nor on $u \in V$, nor on $T$, nor on $T'$.
\end{proof}

A straightforward calculation gives the following.
\begin{lemma}
Let $u \in V$. Then $Au$ is partially differentiable with respect to $T$ twice $(0 \leq T \leq \tau)$, and
\[
\frac{\,\partial Au\,}{\partial T}, \   \frac{\,\partial^2 Au\,}{\partial T^2}  \in C([0,\, \tau] \times [0,\,\hslash\omega_D]).
\]
\end{lemma}

The preceding lemmas imply the following.
\begin{lemma} \quad 
$\displaystyle{ AV \subset V}$.
\end{lemma}

\begin{lemma}\label{lm:setaw}
The set $AV$ is relatively compact.
\end{lemma}

\begin{proof}
Let $u \in V$. Lemma \ref{lm:delta} then implies 
\[
Au(T,\, x)\leq \Delta_2(0)=\frac{\, \hslash\omega_D \,}{\,\sinh\frac{1}{\,U_2\,}\,}.
\]
So the set $AV$ is uniformly bounded. As mentioned in the proof of Lemma \ref{lm:contin}, the $\delta$ does not depend on $u \in V$. Hence the set $AV$ is equicontinuous. The result thus follows from the Ascoli--Arzel$\grave{\mbox{a}}$ theorem.
\end{proof}

\begin{lemma}\label{lm:acontinuous}
The operator $A:\,  V \to V$ is continuous.
\end{lemma}

\begin{proof}
Let $u,\, v \in V$. Then combining a similar discussion to that in the proof of Lemma \ref{lm:gamma} with \eqref{eqn:delta1} gives
\begin{eqnarray}
& & \left| Au(T,\,x)-Av(T,\,x) \right| \nonumber \\
&\leq& U_2 \int_0^{\hslash\omega_D} \frac{1}{\,\left( \xi^2+d^2 \right)^{3/2} \,}
\left\{ \xi^2\tanh \frac{\,\sqrt{\,\xi^2+d^2\,}\,}{2T} + \frac{\, d^2\, \sqrt{\,\xi^2+d^2\,}\,}{\, 2T\,\cosh^2 \frac{\,\sqrt{\,\xi^2+d^2\,}\,}{2T}\,} \right\} \left| u(T,\,\xi)-v(T,\,\xi) \right| \, d\xi \nonumber \\
&\leq& U_2 \int_0^{\hslash\omega_D}  \frac{1}{\,\sqrt{ \xi^2+d^2 } \,} 
\tanh \frac{\,\sqrt{\,\xi^2+d^2\,}\,}{2T} \, d\xi \, \left\| u-v \right\| \nonumber \\
&\leq& \frac{\, U_2\,}{U_1} \int_0^{\hslash\omega_D}  \frac{U_1}{\,\sqrt{ \xi^2+\Delta_1(T)^2 } \,} 
\tanh \frac{\,\sqrt{\,\xi^2+\Delta_1(T)^2\,}\,}{2T} \, d\xi \, \left\| u-v \right\| \nonumber \\
&=& \frac{\, U_2\,}{U_1} \, \left\| u-v \right\|. \nonumber
\end{eqnarray}
Here, $d$ is between $u(T,\,\xi)$ and $v(T,\,\xi)$, and $\| \cdot \|$ denotes the norm of the Banach space $C([0,\, \tau] \times [0,\,\hslash\omega_D])$. The result thus follows.
\end{proof}

We now extend the domain $V$ of our operator $A$ to the closure $\overline{V}$. Let $u \in \overline{V}$. Then there is a sequence $\{ u_n \}_{n=1}^{\infty} \subset V$ satisfying $\| u-u_n \| \to 0$ as $n \to \infty$. A similar discussion to that in the proof of Lemma \ref{lm:acontinuous} gives $\{ Au_n \}_{n=1}^{\infty} \subset V$ is a Cauchy sequence, and hence there is an $Au \in \overline{V}$  satisfying $\| Au-Au_n \| \to 0$ as $n \to \infty$. Note that $Au \in \overline{V}$ does not depend on how to choose the sequence $\{ u_n \}_{n=1}^{\infty} \subset V$. We thus have the following.

\begin{lemma}
$A:\,  \overline{V} \to \overline{V}$.
\end{lemma}

It is not obvious that $Au$ \   $(u \in \overline{V})$ is expressed as \eqref{eqn:ouroperator}. The next lemma shows this is the case. A similar discussion to that in the proof of Lemma \ref{lm:acontinuous} gives the following.

\begin{lemma}\label{lm:integrali}
Let $u \in \overline{V}$. Then
\[
Au(T,\,x)=\int_0^{\hslash\omega_D}
\frac{U(x,\,\xi)\, u(T,\, \xi)}{\,\sqrt{\,\xi^2+u(T,\, \xi)^2\,}\,}\,
\tanh \frac{\,\sqrt{\,\xi^2+u(T,\, \xi)^2\,}\,}{2T}\, d\xi.
\]
\end{lemma}

\begin{proof}
For $u \in \overline{V}$, set
\[
I(T,\,x)=\int_0^{\hslash\omega_D}
\frac{U(x,\,\xi)\, u(T,\, \xi)}{\,\sqrt{\,\xi^2+u(T,\, \xi)^2\,}\,}\,
\tanh \frac{\,\sqrt{\,\xi^2+u(T,\, \xi)^2\,}\,}{2T}\, d\xi, \quad
(T,\,x) \in [0,\, \tau] \times [0,\,\hslash\omega_D]
\]
and let $\{ u_n \}_{n=1}^{\infty} \subset V$ be a sequence satisfying $\| u-u_n \| \to 0$ as $n\to\infty$. Note that the function $(T,\,x) \mapsto I(T,\,x)$ just above is well-defined and continuous. Then
\[
| Au(T,\,x)-I(T,\,x) | \leq | Au(T,\,x)-Au_n(T,\,x) |+ | Au_n(T,\,x)-I(T,\,x) |.
\]
Since $Au_n \to Au$ in the Banach space $C([0,\, \tau] \times [0,\,\hslash\omega_D])$, the first term on the right becomes
\[
 | Au(T,\,x)-Au_n(T,\,x) | \leq  \| Au-Au_n \| \to 0 \quad  (n \to \infty).
\]
A similar discussion to that in the proof of Lemma \ref{lm:acontinuous} gives the second term becomes
\[
 | Au_n(T,\,x)-I(T,\,x) | \leq \frac{\, U_2\,}{U_1} \, \left\| u_n-u \right\| \to 0 \quad  (n \to \infty).
\]
The result thus follows.
\end{proof}

Similar discussions to those in Lemmas \ref{lm:delta} and \ref{lm:gamma} give the following.

\begin{lemma}\label{lm:vvar}
Let $u \in \overline{V}$ and let $\gamma$ be as in \eqref{eqn:gamma}.  Then $\Delta_1(T) \leq Au(T,\,x) \leq \Delta_2(T)$. Moreover, if $T<T'$, then $0 \leq Au(T,\, x)-Au(T',\, x) \leq \gamma (T'-T)$.
\end{lemma}

Lemma \ref{lm:vvar} implies $Au(T,\,x) \leq \Delta_2(0)$ for $u \in \overline{V}$ since the function $\Delta_2$ is strictly decreasing with respect to the temperature $T$. Hence the set $A\overline{V}$ is uniformly bounded. Similar discussions to those in the proofs of Lemmas \ref{lm:contin} and \ref{lm:setaw} give the following.

\begin{lemma}\label{lm:vvarequcon}
Let $u \in \overline{V}$. Then $Au \in C([0,\, \tau] \times [0,\,\hslash\omega_D])$. Moreover, the set $A\overline{V}$ is equicontinuous, and hence the set $A\overline{V}$ is relatively compact.
\end{lemma}

By Lemma \ref{lm:integrali}, a smilar discussion to that in the proof of Lemma \ref{lm:acontinuous} gives the following.

\begin{lemma}\label{lm:avarcontin}
The operator $A:\,  \overline{V} \to \overline{V}$ is continuous.
\end{lemma}

Lemmas \ref{lm:vvarequcon} and \ref{lm:avarcontin} imply the following.

\begin{lemma}\label{lm:acompact}
The operator $A:\,  \overline{V} \to \overline{V}$ is compact.
\end{lemma}

Combining Lemma \ref{lm:acompact} with Lemma \ref{lm:setvvar} and then applying the Schauder fixed-point theorem give the following.

\begin{lemma}
The operator $A:\,  \overline{V} \to \overline{V}$ has at least one fixed point $u_0 \in \overline{V}$, i.e., $\displaystyle{ u_0=Au_0 }$.
\end{lemma}

The uniqueness of the nonzero fixed point of $A:\,  \overline{V} \to \overline{V}$ was pointed out in Theorem \ref{thm:3-1}. Our proof of Theorem \ref{thm:main} is now complete.

\section*{Acknowledgments}
S. Watanabe is supported in part by the JSPS Grant-in-Aid for Scientific Research (C) 24540112. K. Kuriyama is supported by the Yamaguchi University Foundation.

\end{document}